\documentclass{aa}  

\usepackage{graphicx}
\usepackage{txfonts}
\usepackage{color}
\newcommand{\dg}{^{\circ}}
\usepackage{hyperref}
\begin{document}

   \title{Structural axes of sources in the SUMSS survey and their alignments}


   \author{D. Blinov\inst{1,2}\fnmsep\thanks{\email{blinov@ia.forth.gr}}
          \and
          T. Bhattarai\inst{3,4}
          \and
          I. Liodakis\inst{1,5}
          \and
          A. Kumar\inst{2,1}
          \and
          P. Bonfini\inst{6}
          }

   \institute{Institute of Astrophysics, Foundation for Research and Technology - Hellas, Voutes, 70013 Heraklion, Greece
        \and
        Department of Physics, University of Crete, 70013, Heraklion, Greece      
        \and
        Department of Mathematics, Universit\"at T\"ubingen, Auf der Morgenstelle 10, 72076 T\"ubingen, Germany
        \and
        Constructor University, Campus Ring 1, 28759 Bremen, Germany
        \and
        Institute of Astrobiology, University of Crete, 70013, Heraklion, Greece
        \and
        ALMA Sistemi Srl, Guidonia (Rome), 00012, Italy\\
             }

   \date{Received XX; accepted YY}

 
  \abstract
   {Coherent orientations of extragalactic radio sources have been reported on large angular scales, but their statistical significance and susceptibility to observational systematics remain debated. Searches based on very long baseline interferometry (VLBI) also leave the far southern sky comparatively sparsely sampled.}
   {We test for local alignments of kiloparsec-scale radio source structural axes in the Sydney University Molonglo Sky Survey (SUMSS), providing a complementary probe in the undersampled southern sky.}
   {We constructed a catalog of structural position angles for 4087 sources from 629 SUMSS mosaics. A convolutional neural network supplied initial angle estimates, which were corrected for image-to-sky coordinate distortion and visually vetted. We searched for local alignments using a coordinate-invariant parallel transport statistic on HEALPix tessellations, then refined candidate regions with a genetic algorithm. We calibrated the significance of the full search against mock skies that preserve the declination-dependent position angle distribution and undergo the same search procedure.}
   {Several regions show apparently strong local alignment, but none remains statistically significant after accounting for the look-elsewhere effect. The candidate signals are consistent with chance fluctuations in the full search.}
   {We find no statistically significant evidence for alignment of structural axes of radio sources in SUMSS images. The resulting southern sky position angle catalog provides a useful resource for future studies of radio source morphology and population properties.}


   \maketitle
%

\section{Introduction}
\label{sec:introduction}

The cosmological principle assumes that the Universe is statistically homogeneous and isotropic on sufficiently large scales. It underlies the Friedmann–Lema\^itre–Robertson–Walker metric and the standard $\Lambda$ cold dark matter ($\Lambda$CDM) model of cosmology. Within this framework, the matter distribution is expected to approach homogeneity above some transition scale, although the precise value depends on how the transition is defined. Using the fractal dimension, \citet{Yadav2010} derived a theoretical upper limit of $\sim 260\,h^{-1}\,$Mpc for the homogeneity scale in $\Lambda$CDM, while observational estimates from galaxy surveys are typically smaller, of order $70$--$150,h^{-1},$Mpc. A statistically significant excess of coherent structure on scales well beyond this limit, relative to $\Lambda$CDM realizations, would therefore be in tension with the standard cosmological paradigm.

Over the past two decades a growing number of observations have been reported that are difficult to reconcile with this expectation \citep{Aluri2023}. These include structures in the galaxy and quasar distribution that extend over several hundred megaparsecs to a few gigaparsecs. Examples are the Huge Large Quasar Group \citep{Clowes2013} and the Giant GRB Ring, identified through the redshifts of $\gamma$-ray bursts \citep{Balazs2015}. Another example is a population of correlated large quasar groups whose long axes appear to be preferentially aligned or orthogonal to one another across scales of order a gigaparsec \citep{Friday2022}. More recently, an extended search using an enlarged sample of gamma-ray bursts has argued that the Hercules-Corona Borealis Great Wall extends over an even broader redshift range than previously reported, $0.33 \lesssim z \lesssim 2.43$ \citep{Horvath2025}. A broad review of these and related anomalies is given by \citet{Aluri2023}. Such claims remain contested, however. For example, \citet{Fujii2024} reanalyzed the orientations of large quasar groups in three dimensions, using rotationally invariant statistics on the celestial sphere, and found no evidence that their axes are distributed non-randomly. They also showed that most of the quasar groups are not significantly overdense relative to the field. 

A particularly intriguing subclass of these anomalies concerns coherent, large-scale alignments in the orientation of active galactic nuclei (AGN) and their associated structures. \citet{Hutsemekers1998} first reported that the linear polarization vectors of quasars are coherently oriented over regions of the sky several hundred megaparsecs to a gigaparsec across; this result was confirmed and extended using larger, independent samples and coordinate-invariant statistical tests \citep{Hutsemekers2001,Jain2004,Hutsemekers2005}. Analogous alignments were subsequently reported in the radio band, where the polarization angles of quasars in the JVAS/CLASS surveys were found to be coherently oriented on scales of order $100$\,Mpc \citep{Tiwari2013,Pelgrims2015}. Strikingly, the polarization vectors of quasars belonging to large quasar groups tend to be oriented parallel or perpendicular to the major axis of their host group, suggesting a direct physical link between the alignment phenomenon and the large-scale structure traced by the quasars themselves \citep{Hutsemekers2014,Pelgrims2016}. Several theoretical mechanisms have been proposed to explain these correlations, ranging from cosmic-string-induced alignment of quasar spin vectors within specific large quasar groups \citep{Slagter2021,Slagter2022} to a single correlated background magnetic field capable of reproducing the alignment signal simultaneously at cluster, $100$\,Mpc, and gigaparsec scales \citep{Tiwari2022}.

The polarization alignment phenomenon raises a related question: are the structural axes of radio AGN, defined by the position angles of their radio jets and lobes, coherently oriented across the sky? Radio galaxies, whether compact or extended, are classically divided into Fanaroff-Riley types I and II according to their radio morphology and luminosity \citep{Fanaroff1974}, and the position angle of the radio axis is thought to trace the spin axis of the central supermassive black hole. Using deep Giant Metrewave Radio Telescope imaging of the ELAIS-N1 field, \citet{Taylor2016} reported that radio-galaxy position angles are aligned along a preferred direction on scales of about a degree. A larger sample of sources from FIRST and the TIFR GMRT Sky Survey, compiled through citizen science approach, yielded a $3.2\sigma$ alignment signal on similar angular scales. However, a search in a second, sparser sample did not reproduce the effect \citep{Contigiani2017}. Subsequent analyses of FIRST radio-galaxy position angles found further evidence for alignment on scales below about a degree, but could not exclude the possibility that the signal originates in an as yet unidentified systematic bias rather than a genuine physical effect \citep{Panwar2020}. \citet{Osinga2020} likewise reported a significant departure from a uniform distribution of radio source position angles in the deep LOFAR Two-meter Sky Survey. \citet{Lao2024} found similar alignment signals in a catalog of Fanaroff–Riley type II radio galaxies from FIRST, which was compiled using machine learning. However, they could not fully separate these signals from possible survey systematics. Other work casts doubt on whether such alignments are real. \citet{Simonte2023} reanalyzed the same ELAIS-N1 field studied by \citet{Taylor2016}. They used a much larger sample with both photometric and spectroscopic redshifts, and applied five independent statistical tests. They found no significant two- or three-dimensional alignment of radio-galaxy position angles on any of the scales probed. The literature on kiloparsec-scale jet alignments is therefore markedly inconsistent, with claimed detections, non-detections, and explicit warnings about systematic contamination coexisting for overlapping samples and sky areas.

Most recently, several studies have sought a direct physical link between AGN jet orientation and the surrounding cosmic web of filaments, clusters, and voids, rather than treating alignment as a purely statistical, sky-projected phenomenon. \citet{Jung2025} combined LOFAR radio observations with a three-dimensional catalog of filaments derived from the Sloan Digital Sky Survey. They found that the optical major axes of massive galaxies hosting radio sources are significantly aligned with the nearest cosmic filament, out to distances of several megaparsecs. Radio jets are typically oriented perpendicular to the optical major axis of their host. However, the jets become markedly \emph{more} random in orientation close to filaments. The authors conclude that any large-scale coherence in jet position angles most likely arises indirectly. It would follow from the alignment of galaxies with filaments and of jets with their host galaxies, rather than from a direct, coherent orientation of the jets themselves. \citet{Mahato2025} carried out a complementary study of giant radio galaxies. These are sources whose projected linear size exceeds $0.7$ Mpc. They found that the jet axes of these galaxies are preferentially oriented at large angles to the local filament spine, at a significance of about $3\sigma$. A smaller control sample of ordinary Fanaroff-Riley type II sources showed no such preference. The authors interpret this as evidence that giant radio galaxies grow by expanding into channels of lower density that run perpendicular to filaments. Finally, \citet{Peirani2025} used a cosmological hydrodynamical simulation that evolves black hole spin consistently with the rest of the physics. The simulation cannot resolve jets directly, so the authors used the direction of the black hole spin as a proxy for the jet axis. They showed that this direction is statistically aligned with the rotation axis of the host galaxy. This gives a physical basis for the idea that jet orientation is not independent of the large-scale environment of the host galaxy.

Taken together, these results paint a picture in which coherent alignments of AGN polarization and structural axes have been repeatedly reported across a wide range of scales, wavelengths, and source populations, yet remain the subject of active debate regarding their statistical robustness, their possible instrumental or selection origin, and their physical connection to the cosmic web. A further limitation of the existing kiloparsec-scale studies is their near total reliance on radio continuum surveys of the northern sky, principally the VLA FIRST survey and, more recently, LOFAR, leaving the southern celestial hemisphere essentially untested for this class of anomaly. The Sydney University Molonglo Sky Survey \citep{Bock1999} provides a catalog of radio continuum sources across the sky south of declination $-30^{\circ}$. Its angular resolution is coarser than that of FIRST, but it still resolves the structure of many extended radio sources. It therefore offers a natural and largely independent dataset for extending the search for alignments of structural axes. It covers a region of the sky not previously probed by this type of analysis. It also makes it possible to assess whether any signal could arise from an instrumental systematic specific to one hemisphere.

In this paper we use the SUMSS catalog to construct a sample of extended radio sources with well-determined structural position angles and search for statistically significant alignments in their orientation across a range of angular scales. We use the statistical framework of \citet{Jain2004} and \citet{Contigiani2017}, which is invariant under coordinate changes and relies on parallel transport. This places our results on the same footing as previous studies of alignments of polarization vectors and structural axes. By extending the search for radio axis alignments to the southern sky with an independent survey, we aim to provide a critical test of whether the alignment phenomenon reported in the northern hemisphere reflects a genuine, large-scale physical effect.


\section{Data and data analysis}
\label{sec:data}

\subsection{SUMSS images}
\label{sec:data_images}

We use the images of the Sydney University Molonglo Sky Survey \citep[SUMSS;][]{Bock1999}, a survey of the sky south of declination $-30^{\circ}$ carried out at 843\,MHz with the Molonglo Observatory Synthesis Telescope. We collected all 629 extragalactic mosaics.
Each mosaic covers $4.3^{\circ}\times4.3^{\circ}$, with a small overlap between neighboring mosaics, and is stored in units of Jy\,beam$^{-1}$. All sources in a mosaic are convolved to a common synthesized beam of $45''\times45''\,\mathrm{cosec}|\delta_c|$ (beam position angle $0^{\circ}$), where $\delta_c$ is the declination of the mosaic center. The pixel scale is $11''$ along right ascension and $11''\,\mathrm{cosec}|\delta_c|$ along declination, i.e.\ it follows the elongation of the beam, so that the beam is circular ($\approx4.1$ pixels FWHM) on the pixel grid of every mosaic. All image-plane operations described below are performed on this grid.

\subsection{Source extraction}
\label{sec:data_extraction}

To obtain a catalog of radio sources we ran the Python Blob Detector and Source Finder \citep[PyBDSF,][]{Mohan2015} on each of the 629 mosaics separately. We found the published SUMSS catalog to be incomplete for our purposes and therefore used our own extraction. PyBDSF estimates the background mean and rms maps of the image, identifies islands of emission above a threshold, fits each island with one or several two-dimensional Gaussians, and groups the Gaussians into sources. We set the observing frequency to 843\,MHz and otherwise used the default parameters. For each mosaic we retained the list of fitted Gaussian components, which contains their sky and pixel coordinates, peak and total flux densities, fitted major and minor axes and position angle, and the parameters of the mosaic beam, and we stored these lists in a relational database. The resulting catalog contains $355855$ Gaussian components. Extended sources, in particular radio galaxies with jets and lobes, are frequently split by PyBDSF into several components. This is taken into account as described in Sect.~\ref{sec:data_application}.

\subsection{Training sample of jet position angles}
\label{sec:data_training}

We formally define the jet direction as the position angle (PA) of the major axis of the radio source, measured from north through east in the range $0\dg$--$360\dg$, relative to the central (brighter) component where one is present. Given the relatively low angular resolution of SUMSS, it is often difficult to determine unambiguously whether an elongated radio structure traces a jet, or even whether the source is an AGN at all. However, at the observing frequency and flux density range of SUMSS, the majority of sources are indeed AGN, and their elongation is expected to reflect the jet direction. The major axis therefore traces the jet in most cases. For brevity, we refer to this quantity as the jet PA throughout. In the alignment analysis, PAs are taken modulo $180\dg$, since the radio-source axis is undirected.

To build a training set for a machine-learning algorithm we visually inspected images of components from the catalog and selected 203 sources with a clear jet or a clear elongated structure originating from a central component. For each of them we made a subimage of $\approx40\times40$ pixels ($\approx7.3'$ in right ascension and $7.3'\,\mathrm{cosec}|\delta|$ in declination) centered on the cataloged position. The PA was measured in the image plane with one of two methods, chosen according to the morphology of the source.

The first method is used for a central source with a clear jet on one side. We define a strip that starts at the center of the subimage, is one pixel wide, and points to the northern edge of the frame, with a typical length of 10 pixels. The subimage is rotated in steps of $0.5^{\circ}$ (720 steps, with interpolation), and for every step we sum the pixel values inside the strip. This gives the azimuthal distribution of the surface brightness around the core. A jet is considered detected if the maximum of this distribution exceeds its mean by more than $2.4$ times its standard deviation, and the PA is the rotation angle at which the maximum is reached.

The second method is used when the first one does not detect a jet, typically for sources that consist of a compact core and a separate, roughly circular blob. We mask the central component (a circle of 5 pixels radius) and find the brightest pixel in the remaining subimage. The PA is then the direction from the center to this pixel. The blob is accepted only if its brightest pixel exceeds the mean of the masked subimage by more than $3\sigma$. Additionally, a circular Gaussian with a constant background, fitted to the masked subimage, must converge to a centroid within 10 pixels of the center, with a width $1<\sigma<25$ pixels and an amplitude exceeding the background level.

Because of the variety of morphologies, the measurements were not fully automatic: every subimage was examined by eye, and the strip length (and, when needed, the method) was adjusted until the direction found agreed with the visible jet. This sample serves only to train and test the neural network described below.

\subsection{Convolutional neural network for jet directions}
\label{sec:data_cnn}

Measuring the jet direction with the procedure above requires case-by-case tuning and cannot be applied to the hundreds of thousands of components in our catalog. We therefore trained a convolutional neural network (CNN) to predict the jet PA directly from an image. We treat this as a regression problem, with the PA divided by $360^{\circ}$ as the target. The CNN was implemented with the Keras interface of TensorFlow.

\paragraph{Input and data augmentation.} The input is a $40\times40$ pixel subimage centered on the brightest pixel within a few pixels of the cataloged position of the component, and normalized to the maximum pixel value. The 203 measured sources are too few to train a network, so we augmented the sample using transformations for which the PA changes in a known way. For every source we used (i) the original subimage, (ii) its mirror images with respect to the vertical axis (east--west flip, $\mathrm{PA}\rightarrow360^{\circ}-\mathrm{PA}$) and the horizontal axis (north--south flip, $\mathrm{PA}\rightarrow180^{\circ}-\mathrm{PA}$), and (iii) 25 rotations by angles drawn at random between $1^{\circ}$ and $358^{\circ}$ ($\mathrm{PA}\rightarrow\mathrm{PA}+\phi$ modulo $360^{\circ}$), each of which was again mirrored in the two ways. This gives 78 images per source and $15,834$ images in total, with an almost uniform distribution of PA.

\paragraph{Architecture and training.} The architecture is summarized in Table~\ref{tab:cnn}. It consists of two convolutional blocks, each containing two $3\times3$ convolutional layers with 32 filters followed by $2\times2$ max pooling with stride 2, followed by two fully connected layers with 32 neurons each and a single linear output neuron. Rectified linear unit (ReLU) activations are used in all layers except the output. To limit over-fitting we apply a dropout with a rate of 0.2 after the first dense layer and an L1L2 penalty (with both coefficients equal to 0.8) to the weights of this layer. The network has $\approx8\times10^{4}$ trainable parameters. We used the L1 loss (the sum of absolute differences between predicted and true values) and the Adam optimizer with an initial learning rate of $10^{-4}$ and a batch size of 64. The network was trained for 500 epochs. After epoch 200 the learning rate was reduced by 5\% per epoch.

\begin{table}
\caption{Architecture of the CNN used to predict jet position angles.}
\label{tab:cnn}
\centering
\begin{tabular}{lll}
\hline\hline
Layer & Output shape & Details \\
\hline
Input & $40\times40\times1$ & \\
Convolution ($\times2$) & $36\times36\times32$ & $3\times3$ kernel, ReLU \\
Max pooling & $18\times18\times32$ & $2\times2$, stride 2 \\
Convolution ($\times2$) & $14\times14\times32$ & $3\times3$ kernel, ReLU \\
Max pooling & $7\times7\times32$ & $2\times2$, stride 2 \\
Flatten & 1568 & \\
Dense & 32 & ReLU, L1L2 regul. \\
Dropout & 32 & rate 0.2 \\
Dense & 32 & ReLU \\
Dense (output) & 1 & linear \\
\hline
\end{tabular}
\end{table}

\paragraph{Performance.} The augmented sample was randomly divided into a training set (80\%, $12667$ images) and a test set (20\%, $3167$ images). On the test set the mean absolute difference between the predicted and the measured PA is $15.9^{\circ}$ (computed without taking into account the $360^{\circ}$ wrap-around), and individual predictions can deviate from the measured value by much more than this. The CNN is therefore not used as a final measurement: it only proposes a jet direction, and every proposal is verified visually as described in Sect.~\ref{sec:data_inspection}.

\subsection{Application to the whole catalog}
\label{sec:data_application}

Before applying the CNN we removed the components that cannot have a measurable jet direction. For every component of the catalog we cut a $40\times40$ pixel subimage and rejected it if
(i) it lies closer than 25 pixels to the edge of the mosaic ($28339$ components);
(ii) the brightness of the central pixel is below 60\,mJy\,beam$^{-1}$ ($268663$ components); or
(iii) the shape of the emission is not elongated. For this criterion we found the isophote at twice the standard deviation of the pixel values in the subimage which lies closest to the center, required that it encloses the center, fitted it with an ellipse by least squares, and rejected components with an eccentricity $e<0.6$ or $e\geq0.98$ ($45217$ components), i.e.\ round sources and nearly one-dimensional features.
The remaining $13636$ components were passed to the CNN, which assigned a jet direction to each of them.

\begin{figure*}
   \centering
   \includegraphics[width=0.9\textwidth]{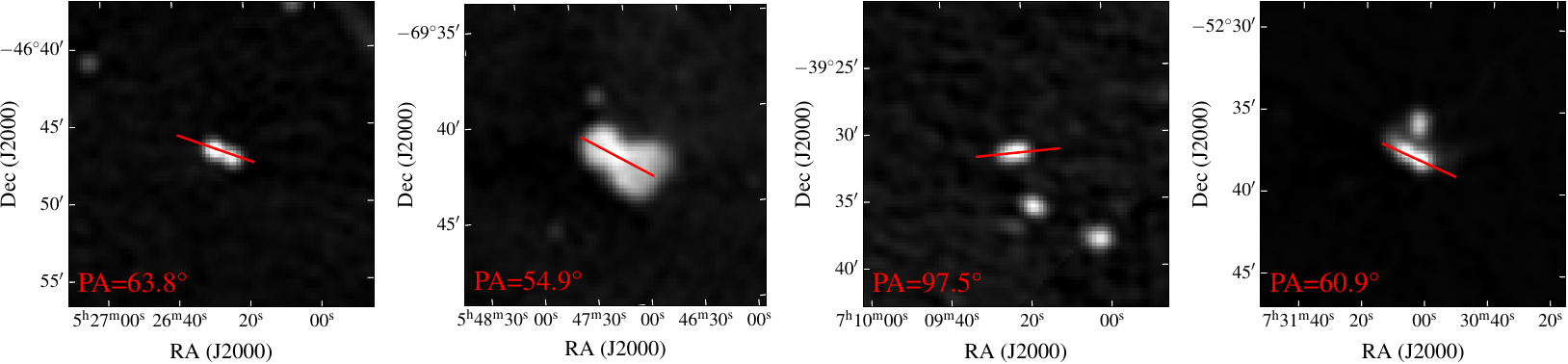}
   \caption{Four examples of structural axis position angle detection. The underlying images are $40\times40$ pixel subimages of original SUMSS tiles used by the CNN.}
              \label{fig:jet_example}
\end{figure*}

As mentioned above, PyBDSF often fragments an extended source into several Gaussian components. Their nearest-neighbor separations show an excess at $\lesssim50''$ compared to randomly distributed sources, indicating components of the same source. We therefore grouped components that lie inside the $2\sigma$ isophote of a brighter neighbor and are closer than $200''$ to it, and displayed them on one common image. This reduced the $13636$ components to $6965$ distinct images. On each image the CNN direction of every component was drawn as an arrow starting at that component, with a different color for each component.

\subsection{Visual inspection and correction for distortions}
\label{sec:data_inspection}

Every image was examined visually through a dedicated web interface. For each component shown on an image the inspector marked whether the arrow follows the axis of a real jet or lobe structure, and could mark components that belong to the same physical source. The inspection was carried out by four members of the team. Most images were examined by a single inspector. A component was retained if all inspectors who examined it judged its direction to be correct. This selected $9511$ of the $13636$ components. For sources with several accepted components, the PAs were reduced to the same half-plane (differences modulo $180^{\circ}$), combined with a circular mean, and the coordinates were averaged with weights inversely proportional to the positional uncertainties. Components manually identified as sub-components of one larger source were combined in the same way if their PAs agreed to within $30^{\circ}$. The combined catalog contains $4882$ sources with a measured jet direction in 628 mosaics, covering declinations from $-88^{\circ}$ to $-30^{\circ}$. A few examples from this catalog overplotted on top of the $40\times40$ pixel subimages are demonstrated in Fig.~\ref{fig:jet_example}.

Because the SUMSS pixel grid is anisotropic (Sect.~\ref{sec:data_images}), a position angle measured in the image plane, $\mathrm{PA}_{\mathrm{im}}$, generally differs from the true position angle on the sky, $\mathrm{PA}_{\mathrm{sky}}$, unless the jet happens to lie along one of the two pixel axes. Two distinct geometric effects contribute to this offset. The first is the anisotropic pixel scale itself: to first order $\tan\mathrm{PA}_{\mathrm{sky}} = \sin|\delta|\tan\mathrm{PA}_{\mathrm{im}}$, an effect that vanishes at the pole and is largest, reaching around 20 degrees, close to the survey's low-declination boundary $\delta=-30^{\circ}$. The second is meridian convergence: each mosaic's pixel grid is aligned with north and east only at its own reference pixel, and a source elsewhere in the tile sits on a different meridian, whose local north points in a different direction. Because SUMSS tiles have a fixed angular width on the sky ($4.3\dg$) rather than a fixed right-ascension range, this effect is negligible at low declination, where a tile subtends only a few degrees of right ascension. However, it becomes severe near the pole, where the same angular width can subtend tens of degrees of right ascension. For example, only 8 tiles cover the whole ring at $\delta=-88\dg$, compared with 71 at $\delta=-32\dg$. Consequently, a source near the edge of a tile can be rotated by a comparable amount relative to the tile's own reference direction. This second effect is not captured by the leading-order formula above and can dominate over it, so that in practice the largest offsets between $\mathrm{PA}_{\mathrm{im}}$ and $\mathrm{PA}_{\mathrm{sky}}$ are found in the polar tiles rather than at low declination.

Because both effects vary with position within a mosaic, we correct each source individually using the WCS of its own tile rather than applying an analytic formula. For every source we start a 50-pixel vector at its cataloged position, oriented along $\mathrm{PA}_{\mathrm{im}}$, and convert its end point to equatorial coordinates with the \texttt{pixel\_to\_world} method of \texttt{astropy.wcs}; the \texttt{position\_angle} method of \texttt{astropy.coordinates} then gives $\mathrm{PA}_{\mathrm{sky}}$ as the position angle on the sky between the source and this end point. Because this uses the full WCS solution rather than a leading-order approximation, it automatically accounts for the pixel-scale anisotropy, the meridian convergence, and their variation across the mosaic.

\subsection{Final cleaning of the sample}
\label{sec:data_cleaning}

The inspection described in Sect.~\ref{sec:data_inspection} was carried out on cutout images, one source at a time. To check the sample in the context of the full mosaics, we displayed each of the 628 mosaics containing selected sources in \texttt{SAOImage DS9}, with the positions of all selected sources overlaid, and inspected them visually. We removed 13 sources that we judged to be spurious, and identified 52 groups of 2--4 entries (113 in total) in the same mosaic that are components of one physical source missed by the earlier grouping. Each such group was replaced by a single source whose coordinates are the medians of the component coordinates and whose PA is the circular median of the component PAs. This left $4808$ sources.

Because neighboring SUMSS mosaics overlap, sources lying in the overlap regions were extracted, inspected and measured independently in two or more mosaics and therefore appear in the sample more than once. We linked sources from different mosaics separated by less than $2'$, which also covers extended sources whose centroids, after the merging of components, differ between mosaics. The links were made in order of increasing separation, and a link was not made if it would place two sources from the same mosaic in one group, so that close but distinct sources seen in two mosaics are not joined. We found 672 sources present in more than one mosaic (637 in two, 21 in three and 14 in four mosaics, the latter at the corners of the mosaic grid). Each group was replaced by a single entry with the mean coordinates and the circular mean of the axial PAs (computed modulo $180^{\circ}$). The final sample contains $4087$ sources.

\begin{figure}
   \centering
   \includegraphics[width=0.48\textwidth]{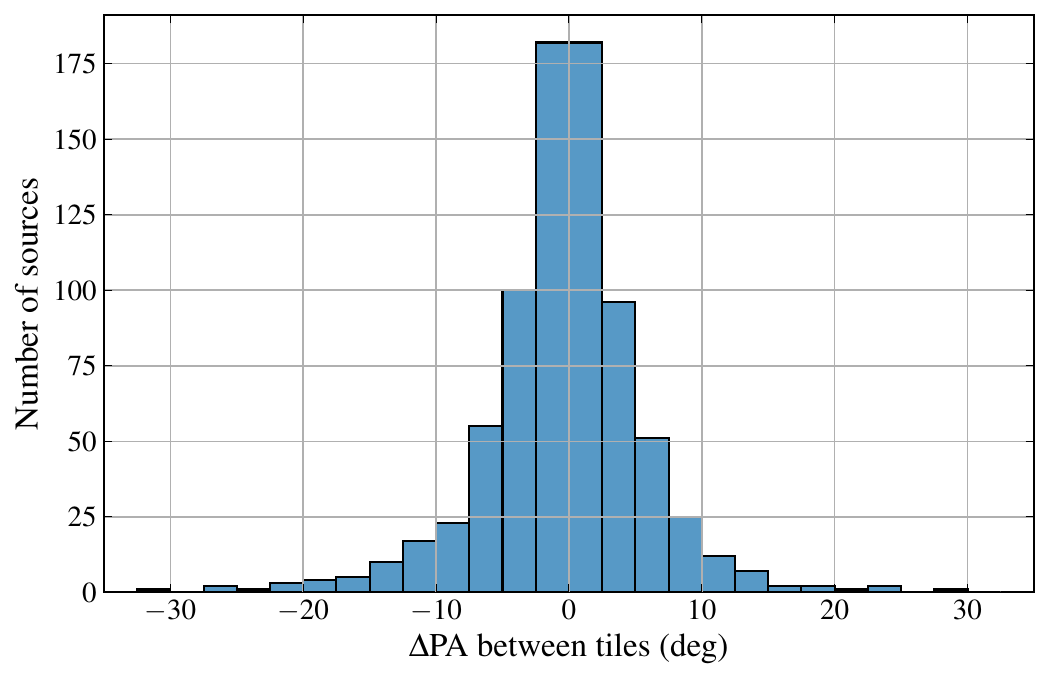}
   \caption{Distribution of the difference between position angles of the same source found in two overlapping tiles.}
              \label{fig:PAdiff}
\end{figure}

The duplicates provide a direct estimate of the uncertainty of our jet directions, since each of them is measured on different mosaics with different pixel grids and WCS solutions, independently extracted by PyBDSF and independently inspected. The distribution of the PA differences between the 784 pairs of mosaics in which the same source was measured is shown in Fig.~\ref{fig:PAdiff}. It is centered at zero (mean $-0.4\dg$) with a standard deviation of $6.0\dg$. Among the pairs 91\% agree within $10\dg$ and only one pair differs by more than $30\dg$. The distribution has wider wings than a Gaussian: its core, estimated from the median absolute deviation, corresponds to a standard deviation of $4.1\dg$. Assuming that both measurements of a pair have the same error, the uncertainty of an individual PA is $6.0\dg/\sqrt{2}\approx4\dg$ (or $\approx3\dg$ for the core of the distribution). The duplicates also test the correction from the image-plane to the sky PA (Sect.~\ref{sec:data_inspection}), because the pixel grids of neighboring mosaics are oriented differently at the position of the same source: before the correction the PA differences of the same pairs have a standard deviation of $15.8\dg$ and only 68\% of them agree to within $10\dg$.

\begin{figure}
   \centering
   \includegraphics[width=0.93\hsize]{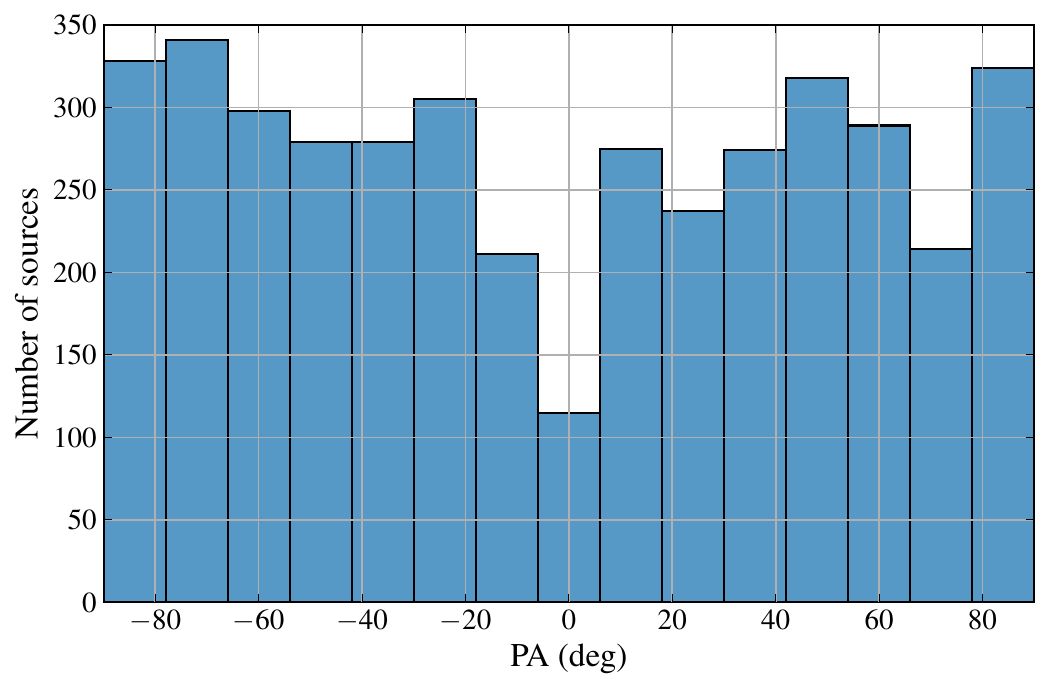}
      \caption{Distribution of position angle of among 4087 sample sources.
              }
         \label{fig:pa_dist}
\end{figure}

Distribution of the structural axis PA in the final sample, reduced to the range $0\dg$--$180\dg$, is demonstrated in Fig.~\ref{fig:pa_dist}. This distribution is not uniform: sources with PA between $[-90\dg, -70\dg]$ and $[70\dg, 90\dg]$ are about 1.5 times more numerous than those between $-20\dg$ and $20\dg$. We attribute this to the anisotropic angular resolution of the survey and take it into account when estimating the significance of jets alignments in the next section.

\section{Search for alignments and significance estimation}
\label{sec:significance}

\subsection{Alignment statistic and parallel transport}
\label{sec:sig_statistic}

\begin{figure}
   \centering
   \includegraphics[width=0.49\textwidth]{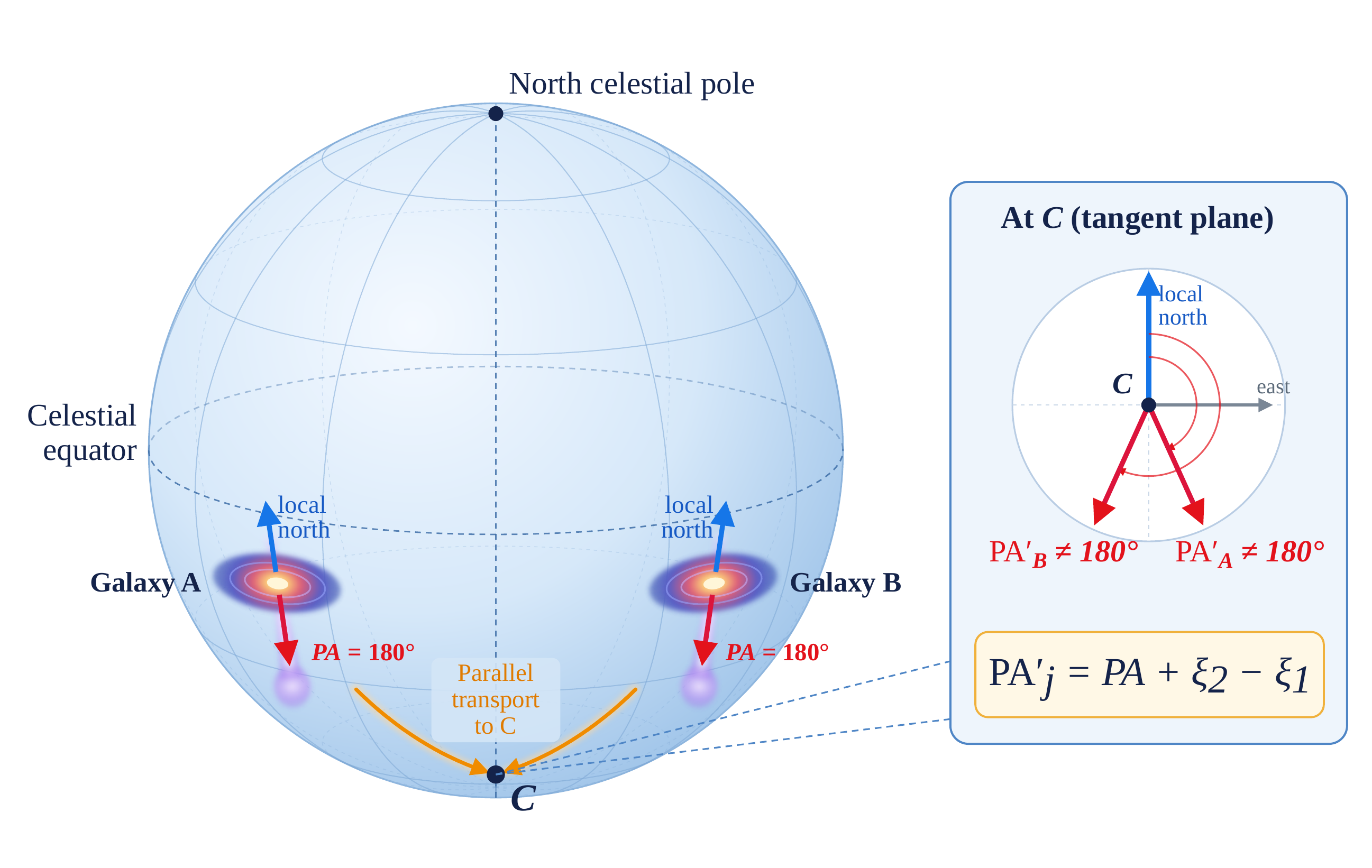}
   \caption{Illustration of the geometrical effect of the local meridian on jet position angles (PAs). Two jets can have the same numerical PA while their projected directions on the celestial sphere are not aligned, because PAs are measured relative to the local meridian at each source position. Parallel transport to a common reference point is therefore required before comparing PAs measured at different sky positions.}
              \label{fig:par_transp}%
\end{figure}

A jet's position angle is measured relative to the local meridian at the position of its host source: it is the angle, at that source, between the direction to the north celestial pole and the direction of the jet. Because the local meridian points in a different direction at every point on the sky, two sources with the same numerical PA are not necessarily aligned with each other in any physical sense. For example, two sources on opposite sides of the north celestial pole can each have $\mathrm{PA}=180\dg$, with both jets pointing directly away from the pole along their own local meridians, while the jets could be in fact perpendicular to one another in space. Figure~\ref{fig:par_transp} illustrates this geometrical effect. PAs measured at different points on the sky therefore cannot be compared, combined, or averaged directly. Each PA must first be referred to a common point before any statistic is computed from them.

We do this with the coordinate-invariant construction of \citet{Jain2004}, as applied to radio-jet position angles by \citet{Contigiani2017} and \citet{Mandarakas2021}. A PA defines a unit tangent vector on the celestial sphere at its source's position. To compare a PA measured elsewhere, we parallel-transport the corresponding vector along the great circle connecting the two points. Unlike in flat space, moving a vector across a curved surface changes its orientation relative to a fixed external reference, which here is local north, even when the vector is kept geometrically unchanged along the path. This path-dependent rotation is precisely what must be corrected for when comparing PAs. For a source at position $j$, transported to a reference point $\mathbf{c}$, the corrected PA is
\begin{equation}
  \mathrm{PA}_j' = \mathrm{PA}_j + \xi_2(\mathbf{c},j) - \xi_1(j,\mathbf{c}),
  \label{eq:parallel_transport}
\end{equation}
where $\xi_1(j,\mathbf{c})$ is the position angle, measured at $j$, of the great circle toward $\mathbf{c}$, and $\xi_2(\mathbf{c},j)$ is the position angle, measured at $\mathbf{c}$, of that same great circle back toward $j$ (both computed exactly on the sphere, \texttt{astropy.coordinates.SkyCoord.position\_angle}); the two differ precisely because local north rotates between the two points. $\mathrm{PA}_j'$ is the orientation source $j$'s jet would appear to have if the source were instead located at $\mathbf{c}$, and it is this transported quantity that can be meaningfully compared between different sources.

For a group of $N$ sources, all transported to a common reference point $\mathbf{c}$ (the center of the HEALPix pixel or candidate polygon under consideration, Sects.~\ref{sec:sig_healpix}--\ref{sec:sig_ga}), we measure their coherence with the resultant length of the doubled transported PAs \citep{Jain2004}:
\begin{equation}
  S = \frac{1}{N}\left|\sum_{j=1}^{N} \exp\left(2i\,\mathrm{PA}_j'\right)\right|.
  \label{eq:S_stat}
\end{equation}
The factor of two maps the axial range $[0\dg,180\dg)$ onto the full circle $[0\dg,360\dg)$, needed because a jet axis, unlike a vector, has no preferred direction along itself. $S=1$ if every $\mathrm{PA}_j'$ is identical (perfect alignment) and $S\to0$ for an isotropic distribution of $N\to\infty$ sources (no alignment); values in between indicate partial coherence, with larger $S$ corresponding to tighter alignment.

\subsection{Local significance: HEALPix search}
\label{sec:sig_healpix}

We first search for alignments on a fixed angular grid. We pixelize the sky with \textsc{HEALPix} \citep{Gorski2005} at five resolutions, $N_{\rm side}\in\{2,4,8,16,32\}$, corresponding to pixels of characteristic angular size $29\fdg3$, $14\fdg7$, $7\fdg3$, $3\fdg7$ and $1\fdg8$ respectively.
For every pixel containing at least 5 sources, we compute the observed statistic $S_{\rm obs}$ (Eq.~\ref{eq:S_stat}, transported to the pixel center) and compare it with the distribution of $S$ expected under the null hypothesis of no alignment, estimated from $N_{\rm MC}=1000$ Monte Carlo realizations of the same group of sources. We use two null models for the PA of an individual source:
\begin{itemize}
  \item \emph{Shuffling}: PAs are drawn, without replacement, from the real PAs of all sources in the same declination band as the pixel (the band spanned by the \textsc{HEALPix} ring the pixel belongs to);
  \item \emph{Generation}: PAs are drawn from a von Mises distribution fitted to the doubled real PAs in that same declination band.
\end{itemize}
Both null models are tied to the local declination band, rather than drawn from a single all-sky distribution, because the observed PA distribution is not uniform (Sect.~\ref{sec:data_cleaning}): it varies with declination as a result of the survey's anisotropic angular resolution. Using the real, local PA distribution as the null model ensures that this known systematic effect cannot by itself produce a spurious alignment detection. Although the null distribution for the von Mises model could be treated analytically under suitable assumptions, we use Monte Carlo simulations to estimate the distribution of the statistic for the actual source configurations and analysis procedure. This approach also provides a consistent framework for comparing the two null models.

For each pixel, we convert the offset between $S_{\rm obs}$ and the Monte Carlo distribution into a one-sided significance level,
\begin{equation}
SL = 1-\Phi\!\left(\frac{S_{\rm obs}-\overline{S}_{\rm MC}}{\sigma_{\rm MC}}\right),
\end{equation}
where $\overline{S}_{\rm MC}$ and $\sigma_{\rm MC}$ are the mean and standard deviation of the Monte Carlo distribution, which we approximate as Gaussian \citep{Simonte2023}, and $\Phi$ is the cumulative distribution function of the standard normal distribution. We quote this one-sided $SL$ as an equivalent number of standard deviations $n$ through
\begin{equation}
SL = 1-\mathrm{erf}\!\left(n/\sqrt{2}\right),
\end{equation}
where $\mathrm{erf}$ is the error function. This expresses the one-sided $SL$ on the same numerical scale as the familiar two-sided Gaussian tail probabilities (e.g.\ $2\,\sigma\simeq4.6\,\%$, $3\,\sigma\simeq0.27\,\%$). As a result, $n$ is somewhat larger than the raw offset $(S_{\rm obs}-\overline{S}_{\rm MC})/\sigma_{\rm MC}$. For example, $n=3.0\,\sigma$ corresponds to a raw offset of about $2.78\,\sigma_{\rm MC}$. Pixels reaching $n\ge3.0\,\sigma$ in either null model are kept as seeds for the refinement described next. On the final sample of $4087$ sources, this selects seven pixels in the shuffling null model, at $N_{\rm side}=4$, $8$ and $16$, and none at $N_{\rm side}=2$ or $32$. The generation null model selects five of the same seven pixels. The remaining two fall just short of the threshold under the generation null model ($2.94\,\sigma$ and $2.97\,\sigma$, compared with $3.05\,\sigma$ and $3.11\,\sigma$ in the shuffling null model).

\subsection{Refining candidate regions: a genetic-algorithm polygon search}
\label{sec:sig_ga}

A fixed \textsc{HEALPix} pixel is a poor match to the true shape of any physical alignment region: its boundary is set by the pixelization, not by the data. A genuine region of coherent PAs may be split across pixel edges or only partly fill a pixel.  We therefore refine every seed pixel found above by searching within a circle of a radius of $1.2$ times the pixel's angular size centered on the pixel. Within this circle, we search for the convex polygon of sources with the highest alignment significance, allowing an arbitrary shape and number of vertices. A convex region can contain anywhere from a handful to several hundred sources within the search circle. Selecting the sources that define the region is therefore a combinatorial optimization problem. Exhaustive search becomes impractical once a circle contains more than a few dozen sources. We therefore solve it with a genetic algorithm and retain only the most significant polygon found for each seed pixel, so that the real data and every mock sky (Sect.~\ref{sec:sig_global}) are searched using the identical procedure and are directly comparable pixel by pixel.

A candidate solution ("chromosome") is encoded as a binary string over the sources inside the search circle, indicating which sources are candidate polygon vertices. Every chromosome is deterministically repaired to a valid convex polygon by replacing it with the convex hull of its selected sources, an operation equivalent to a Graham scan. Thus, a chromosome can never produce a self-intersecting or non-convex shape, so no separate feasibility penalty is needed. Every source inside the search circle that falls within the resulting hull is counted as a member of the candidate region. The fitness of a chromosome is the alignment significance (Sect.~\ref{sec:sig_statistic}) of its member sources, transported to the polygon's centroid. The population (60 chromosomes) is evolved for 80 generations using tournament selection, uniform crossover, bit-flip mutation, and elitism. Half of the initial population is seeded as the $k$ nearest neighbors of a randomly chosen source, which biases the search towards plausible, spatially compact candidate regions from the start. To keep the search affordable, the fitness of every chromosome is evaluated at a reduced $N_{\rm MC}=500$ in each generation. The genome with the highest significance over the whole run is then re-evaluated once at the much higher $N_{\rm MC}=2\times10^{4}$. The final high-precision value is the one reported throughout.

\begin{figure}
   \centering
   \includegraphics[width=0.40\textwidth]{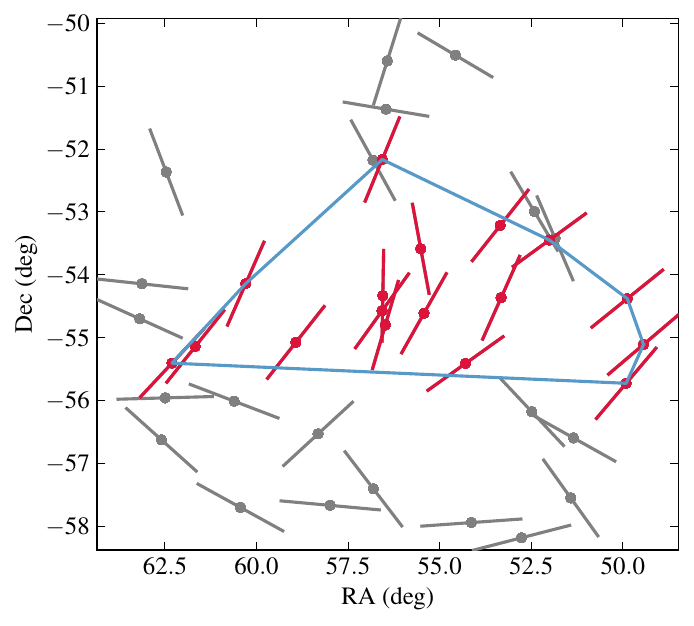}
   \caption{Directions of jets in the most significant candidate region (shuffling null model, $N_{\rm side}=16$). Member sources (the 17 enclosed by the polygon) and their structural axis directions are shown in red, other sources in the region -- in gray, and the polygon boundary found by the genetic algorithm -- in blue.}
              \label{fig:most_sig}%
\end{figure}
\begin{figure}
   \centering
   \includegraphics[width=0.49\textwidth]{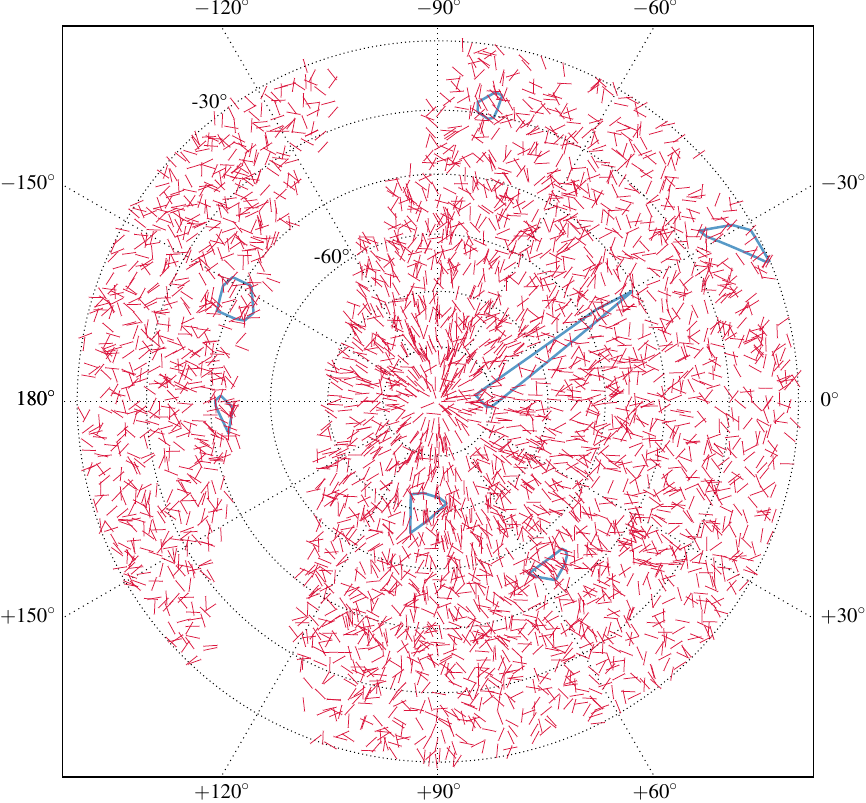}
   \caption{SUMSS survey area in a south polar stereographic projection. The found jet directions are shown by red line segments. The Galactic plane avoidance zone is intrinsic to the SUMMS extragalactic survey. The 7 candidate regions listed in Table~\ref{table:regions} are outlined in blue. The apparent alignment of sources along the meridians near the pole is a projection effect.}
              \label{fig:sky_data}%
\end{figure}

Applied to the seed pixels of Sect.~\ref{sec:sig_healpix}, this search returns 7 regions in the shuffling null model and 5 in the generation null model (one region per seed pixel). Table~\ref{table:regions} shows results for the shuffling null model. The most significant region in the shuffling null model is a 17-source, 7-vertex polygon at $5.07\,\sigma$, around $(\rm{RA},\rm{DEC})\approx(54\fdg3,-54\fdg3)$. This most significant (shuffling null model) polygon is shown in Fig.~\ref{fig:most_sig}, while all 7 most significant shuffling-model polygons, one per seed pixel, are shown in Fig.~\ref{fig:sky_data}, overlaid on the sample's distribution on the sky. In fact, both null models agree on the same two regions as the most significant overall, only with their significance values varying slightly between the null models. These variations are likely due to Monte Carlo noise rather than indicating any disagreement between the two null models.

\subsection{Global significance from mock skies}
\label{sec:sig_global}

A significance of $5\,\sigma$ or more may appear highly significant on its own. However, the search involves a very large number of statistical trials. The \textsc{HEALPix} screening tests thousands of pixels at five resolutions. For each pixel that passes the significance condition, the genetic algorithm then explores a large number of possible convex regions. Even in the absence of genuine alignment, the extensive search can identify apparently significant local features purely by chance. This is the well-known look-elsewhere effect. The quoted local significances (Sect.~\ref{sec:sig_ga}) therefore cannot be interpreted as is, and must be recalibrated against the significance this same search procedure reaches on skies where, by construction, there is nothing to find.

We generated $50000$ such mock skies. For each mock sky, the real sources are grouped into six declination bands ($10\dg$ wide, $-90\dg$ to $-30\dg$). The PAs within each band are randomly shuffled among its sources. The sources themselves are then re-scattered uniformly in area within the same band, excluding Galactic plane exclusion zone, as in the real sample. This preserves, by construction, the real declination-dependent PA distribution discussed in Sect.~\ref{sec:data_cleaning}. It also preserves the anisotropic-resolution effect already included in the null models of Sect.~\ref{sec:sig_healpix}. At the same time, it removes any genuine association between a source's position and its PA. We then applied the identical two-stage search described in Sects.~\ref{sec:sig_healpix}--\ref{sec:sig_ga} to every mock sky. This included the \textsc{HEALPix} search at $3.0\,\sigma$ and the genetic-algorithm refinement with the same population size, number of generations, and other parameters used for the real data. As for the real data, we retained only the single most significant region found around each seed pixel. All $50000$ mock skies have been processed with both null models.

These mock runs let us check how severe the trials factor actually is: of the $50000$ mock skies with no alignment, $96.8\,\%$ (shuffling null model) and $95.1\,\%$ (generation null model) produce at least one \textsc{HEALPix} pixel reaching our $3.0\,\sigma$ search threshold, with an average of $3.8$ and $3.3$ such seed pixels per mock sky, respectively. For each real candidate region, we count the mock skies whose most significant result is at least as significant as the real detection. The global false-alarm probability is then calculated as $p=(n+1)/(N+1)$ \citep[e.g.][]{Davison1997}, where $n$ is the number of such mock skies and $N$ is the total number of mock skies. We convert this probability to a global, trials-corrected significance using the same convention as Sect.~\ref{sec:sig_healpix}.

\subsection{Results}
\label{sec:sig_results}

Once the trials factor is accounted for in this way, none of the candidate regions of Sect.~\ref{sec:sig_ga} remain significant in either null model. The most significant local detection in the shuffling null model, the $5.07\,\sigma$ region of Fig.~\ref{fig:most_sig}, gives a global significance of only $1.61\,\sigma$ ($p=0.106$). The generation null model gives a consistent picture for the same region, $1.41\,\sigma$ ($p=0.160$). Every other candidate region is even less significant, down to essentially $0\,\sigma$ globally for the least significant ones (Table~\ref{table:regions}). We therefore find no statistically significant evidence for coherent position-angle alignment in this sample: the local significances of Sect.~\ref{sec:sig_ga}, although individually as high as $5.07\,\sigma$, are fully consistent with what this same search routinely produces on skies constructed to contain no genuine alignment. Once the look-elsewhere effect of the search procedure itself is properly calibrated against those skies, the global significances are entirely unremarkable.

\begin{table}
\caption{The 7 candidate alignment regions found in the shuffling null model (Sect.~\ref{sec:sig_ga}), ranked by local significance. $N$ is the number of member sources enclosed by the polygon; (RA,Dec) is its centroid; $R$ is its angular radius (the largest centroid-to-vertex separation). Global significances are from repeating the analysis for the $50000$ mock skies at the same $N_{\rm side}$ (Sect.~\ref{sec:sig_global}).}             
\label{table:regions}      
\centering                          
\begin{tabular}{c c c c c c c}        
\hline\hline                 
 $N_{\rm side}$ & RA  & Dec & $R$ & $N$ & \multicolumn{2}{c}{Significance} \\    
       &     &     &     &  & local    & global \\    
       & deg & deg & deg &  & $\sigma$ & $\sigma$ \\    
\hline                        
  16 &  54.34 & $-$54.34 &  4.70 & 17 & 5.07 & 1.61 \\      
  16 &  95.94 & $-$70.60 &  5.33 & 14 & 5.05 & 1.58 \\
   8 & 332.24 & $-$32.12 &  6.52 & 13 & 4.33 & 0.16 \\
   8 & 206.96 & $-$50.52 &  3.84 & 19 & 4.11 & 0.08 \\
  16 & 280.09 & $-$38.81 &  2.42 & 10 & 4.00 & 0.30 \\
   4 &   1.78 & $-$55.23 &  9.33 & 16 & 3.95 & 0.05 \\
  16 & 178.03 & $-$51.91 &  4.12 & 11 & 3.92 & 0.23 \\
\hline                                   
\end{tabular}
\end{table}

\subsection{Redshift-dependent analysis}
\label{sec:sig_redshift}

All of the analysis above treats the sample purely as projected on the sky, with no information on line-of-sight distance. To test whether restricting the sample by redshift changes this picture, we cross-matched our catalog of $4087$ sources against OCARS \citep{Malkin2018}, Quaia \citep{StoreyFisher2024} and MORX \citep{Flesch2024}, using a matching radius of one SUMSS angular resolution element, $45\arcsec$ (Sect.~\ref{sec:data_images}). We estimated the random coincidence rate of each cross-match empirically, by repeating the same nearest-neighbor match against several copies of our catalog with every source's right ascension shifted by a fixed offset. In each copy, every source keeps its declination, and hence its footprint overlap with each external catalog, but is otherwise at an unrelated sky position. This gives an expected chance-match rate of only $\sim0.2\,\%$ for OCARS and $\sim2.4\,\%$ for MORX, but $\sim16\,\%$ for Quaia, reflecting its much higher source density on the sky. Where more than one catalog matched the same source, we kept a single redshift value in priority OCARS $>$ Quaia $>$ MORX. This ordering reflects the nature of each catalog: OCARS is built from the same kind of compact radio sources as this sample, and Quaia is a purpose-built, high-purity quasar catalog. MORX, by contrast, is the broadest and most heterogeneous of the three. This yielded $644$ sources with a redshift estimate ($99$ via OCARS, $346$ via Quaia, $199$ via MORX). We repeated the \textsc{HEALPix} search of Sect.~\ref{sec:sig_healpix} independently in three redshift slices, $z\in[0,0.1]$, $[0.1,0.5]$ and $[0.5,5.0]$ ($118$, $152$ and $374$ sources respectively). In none of the three slices, under either null model, did any \textsc{HEALPix} pixel reach even a  threshold of $3\,\sigma$ (the most significant pixel found, in any slice or null model, was $2.7\,\sigma$). The genetic-algorithm polygon refinement and mock-sky trials correction of Sects.~\ref{sec:sig_ga}--\ref{sec:sig_global} were therefore not run, since there was no candidate region for them to refine or correct. This agrees with the null result for the full sample. Restricting the sample to a single redshift slice also sharply reduces the number of \textsc{HEALPix} pixels that contain enough sources to be tested. Summed over all five \textsc{HEALPix} resolutions, this number falls from several hundred for the full sample to as few as 15 for the lowest redshift slice. There were therefore far fewer chances for a false positive to arise through the look-elsewhere effect, and also less statistical power to detect a genuine signal.

\section{Summary and conclusions}
\label{sec:conclusions}

We have searched for coherent alignment of the structural axes of radio sources in the Sydney University Molonglo Sky Survey. This radio-continuum survey covers the sky south of $\delta=-30\dg$, providing a southern hemisphere sample. This region has not previously been tested for this class of anomaly.

Starting from the $629$ SUMSS mosaics, we extracted a catalog of radio components with \textsc{PyBDSF} and trained a convolutional neural network, using $203$ sources whose PAs we first measured by hand, to estimate the structural position angle (PA) of a radio source directly from its cutout image. The network's output was corrected for the systematic image-plane-to-sky-plane distortion caused by the survey's anisotropic pixel scale, and every predicted PA was visually vetted individually via a dedicated web interface by four team members. Following this vetting, we merged components of the same physical source, removed spurious detections, and combined sources measured independently in overlapping neighboring mosaics. This yielded a final catalog of 4087 sources with measured structural PAs.

We then searched this catalog for statistically significant alignment of these position angles, using the coordinate-invariant, parallel-transport statistic of \citet{Jain2004}. A \textsc{HEALPix}-based search at five angular resolutions, followed by a genetic-algorithm search that refines every screened pixel into the single most significant convex region of sources around it, identified 7 candidate regions reaching a local significance of up to $5\,\sigma$.

Taken in isolation, such a high local significance would indicate a strong detection. However, our two-stage search is extremely thorough by design. The initial \textsc{HEALPix} search tests thousands of pixels across five resolutions, after which the genetic algorithm explores a vast combinatorial space of candidate regions around each pixel. Because of this exhaustive search, a high local significance is expected to arise by chance alone, even where no genuine alignment exists. We quantified this directly by repeating the identical, two-stage search on $50000$ mock skies built by reshuffling the real position angles within declination bands and re-scattering the sources: $96.8\,\%$ of these mock skies, constructed to contain no alignment whatsoever, nonetheless produce at least one region reaching our $3.0\,\sigma$ threshold purely from the look-elsewhere effect. We then compare the real detections with this empirical distribution instead of an idealized null hypothesis for a single trial. None of the seven candidate regions remains significant. The strongest region has a local significance of $5.07\,\sigma$, but only $1.61\,\sigma$ globally. All other regions have even lower global significances, and several are consistent with zero (Table~\ref{table:regions}).

We therefore find no statistically significant evidence for the coherent alignment of radio-source structural axes in the SUMSS sample. This result is directly relevant to the broader literature debate summarized in Sect.~\ref{sec:introduction}. Several kiloparsec-scale studies in the northern sky have reported local alignment significances similar to ours \citep[e.g.][]{Taylor2016,Osinga2020}. However, other studies caution that such signals cannot be distinguished from systematic or look-elsewhere effects without proper calibration \citep{Panwar2020,Simonte2023,Lao2024}. Using independent data from the southern hemisphere and a distinct search methodology, we demonstrate this pitfall quantitatively: our initial local significance exceeded $5\,\sigma$, but a rigorous trials correction reduced it to an unremarkable $1.6\,\sigma$. Although our result does not determine whether the specific northern-sky claims are genuine, it clearly demonstrates that reporting local significance without accounting for the trials factor is an unreliable practice.

Finally, we release the catalog of $4087$ SUMSS sources with their measured structural position angles, coordinates, and associated uncertainties, together with the mosaic and component identifiers linking each entry back to the original SUMSS images. Beyond the alignment search presented here, our catalog is, to our knowledge, among the largest homogeneous catalogs of kiloparsec-scale radio structural axes in the southern sky. This region remains comparatively under-served by such measurements relative to the northern sky, where FIRST- and LOFAR-based samples are available (Sect.~\ref{sec:introduction}). The catalog should also be useful for other studies. Potential applications include investigating alignments between radio jets and host-galaxy structural axes, training or validating automated jet-orientation methods, and combining the catalog with forthcoming deeper southern-sky radio surveys.

\section{Data availability}
\label{sec:data_availability}

Position angles of structural axes of radio sources in the SUMSS survey found and analyzed in this work will be available upon acceptance of the manuscript through Harvard Dataverse at \url{https://dx.doi.org/TBA}.

\begin{acknowledgements}
D.B. acknowledges support from the European Research Council (ERC) under the European Union Horizon 2020 research and innovation programme under the grant agreement No 771282. T.B. acknowledges funding by DAAD-RISE (Deutscher Akademischer Austauschdienst - Research Internships in Science and Engineering) for funding of the project. I.L. was funded by the European Union ERC-2022-STG - BOOTES - 101076343. We wish to thank the AstroStat Academy for providing training on the analysis methods adopted in this work.
\end{acknowledgements}

%
%
\bibliographystyle{aa}
\bibliography{bibliography.bib}

\end{document}